# Comment on Frauchiger and Renner paper (*Nat. Commun*. **9**, 3711 (2018)): The problem of stopping times

P. B. Lerner[1]

Abstract

The *Gedankenexperiment* advanced by Frauchiger and Renner in their *"Nature"* paper was based on an implicit assumption that one can synchronize stochastic measurement intervals between two non-interacting systems. This hypothesis, the author demonstrates, is equivalent to the complete entanglement of these systems. Consequently, Frauchiger and Renner's postulate Q is meaningless and needs to be narrowed. Accurate reformulation of the postulate Q1 does not entail any paradoxes with measurement. This paper is agnostic concerning particular interpretations of quantum mechanics. Nor does it refer to the "collapse of the wavefunction."

[1] Retired and with Device Consultants, LLC, State College, PA 16804, USA. pblerner18@gmail.com, pblermer@syr.edu.

## Introduction

Frauchiger and Renner, in their paper ([FR2018]), further FR, proposed a *Gedankenexperiment*, generalizing Wigner's "friends" paradox ([Wigner1961] ), which can, in principle, be realized with photon polarizations. Since then, there appeared a large literature ([Bub2018], [Laloe2018], [Yang2018], [Sudbery2019] and [Relaño2019]) and *op. cit*., which claims to resolve or assert the philosophical statements made in the FR paper. This author does not claim that the argument below supports any interpretation of quantum mechanics, nor is he competent to judge.

Crudely speaking, a proposed author's objection to the FR protocol consists of the following observation. In quantum mechanics, the probability distribution expressed through a wavefunction or a density matrix, which obeys Schrödinger or Landau-Bloch equations, respectively, is unconditional. Experimentally, though, we always observe distributions conditional on the (quantum) state of the measurement device. Before the 1980s, there was no experiment where this fine distinction could cause confusion. However, thanks to modern quantum optics, playing fast and loose with the quantum state of the device before or after measurement can sometimes lead to paradoxical conclusions.

The original FR argument is based on an implicit assumption that one can synchronize the clocks between two non-interacting quantum mechanical (QM) systems with arbitrary accuracy, at least in non-relativistic quantum mechanics. As we show below, this assumption is equivalent to the maximum entanglement of the two systems. Because of quantum monogamy, it places severe restrictions on the possibility of things being measured.

The paper is structured as follows. In the first section, we observe that the state of the first observer's room after performing Step 1 of the FR is mixed and must be described by the appropriate density matrix. In the second section, we define a real-valued random measurement process. In Section 3, we build the POVM algebra for the "Friends" paradox. In the fourth and fifth sections, we extend our notion of the random measurement process to the operator-valued measures. In the sixth section, we formulate the necessary correction to the FR Condition Q on the quantum measurement, which they consider self-evident but do not identify a specific physical reason for the breakdown or, more accurately, trivialization of their protocol. Section 7 connects the FR setup with the quantum monogamy principle between the friends' rooms. Section 8 concludes the study with the observation that under the FR protocol, either the particle's quantum state or the quantum state of one of the rooms must remain undefined. The conclusion section discusses the possibility to avoid the pitfalls of the FR setup. The Appendix demonstrates that a literal application of textbook quantum mechanics *does not necessarily* lead to Wigner's friend paradox.

## 1. **Note on Step 1 of FR protocol**

In Step 1 of the FR protocol, the observer, whom we indicate as Bub, in reference/reverence to Jeffrey Bub, receives a quantum state:

$$|\psi>_r = \alpha|h>_p + \beta|t>_p \tag{1}$$

Where $|h>$ and $|t>$ are the orthogonal quantum states traditionally identified with heads and tails of a coin, spin z-directions or photon polarizations, and prepares a state:

$$|\psi>_{\mathcal{B}} = \frac{1}{\sqrt{2}}|\uparrow>_{\mathcal{B}=h} + \frac{1}{\sqrt{2}}|\downarrow>_{\mathcal{B}=t} = \\ \frac{1}{\sqrt{2}} <\uparrow|h>_{\mathcal{B}}\ |F=h>_{\mathcal{B}} + \frac{1}{\sqrt{2}} <\downarrow|t>_{\mathcal{B}}\ |\bar{F}=t>_{\mathcal{B}} \tag{2}$$

The first line indicates that Bub prepares a Bell state, in which $|\uparrow>$ and $|\downarrow>$ are the orthogonal polarization vectors conditional on the observer's measurement of heads or tails. The index $\mathscr{B}$ refers to Bub's Hilbert state $\mathscr{L}_{\mathscr{B}}$ and its scalar product. The second equality indicates that the observer's state (density matrix) is conditional on the outcome of the particle measurement. Note that the observer after this procedure is left in the mixed state; otherwise, his wave function would be (compare to [Rovelli96]):

$$|\psi>_{\mathcal{B}}=\frac{1}{\sqrt{2}}|\uparrow>_{\mathcal{B}=h}+\frac{1}{\sqrt{2}}|\downarrow>_{\mathcal{B}=t}=$$

$$\frac{\tilde{\alpha}}{\sqrt{2}}<\uparrow|h>_{\mathcal{B}}\ |F=h>_{\mathcal{B}}+\frac{\tilde{\beta}}{\sqrt{2}}<\uparrow|h>_{\mathcal{B}}\ |F=t>_{\mathcal{B}}+\frac{\tilde{\alpha}}{\sqrt{2}}<\downarrow|t>_{\mathcal{B}}\ |F=h>_{\mathcal{B}}+\frac{\tilde{\beta}}{\sqrt{2}}<\downarrow|t>_{\mathcal{B}}\ |F=t>_{\mathcal{B}}$$

(3)

where $|\tilde{\alpha}|^2+|\tilde{\beta}|^2=1$.

*Proof*

$$Tr_{\mathcal{B}}[\hat{\rho}^2]=Tr[|\psi>_{\mathcal{B}}<\psi|_{\mathcal{B}}|\psi>_{\mathcal{B}}<\psi|_{\mathcal{B}}]=\frac{1}{2}\begin{pmatrix}|<\uparrow|h>_{\mathcal{B}}|^2 & {}_{\mathcal{B}}<F=h|F=h>_{\mathcal{B}}+\\ |<\downarrow|t>_{\mathcal{B}}|^2 & {}_{\mathcal{B}}<F=t|F=t>_{\mathcal{B}}\end{pmatrix}$$

$$=\frac{1}{2}\begin{pmatrix}|<\uparrow|h>_{\mathcal{B}}|^2 1_{\mathcal{B}}+\\ |<\downarrow|t>_{\mathcal{B}}|^2 1_{\mathcal{B}}\end{pmatrix}\leq 1$$

(4)

Where the equality $Tr_{\mathcal{B}}[\hat{\rho}^2]=1$ is reached if and only if $|<\uparrow|h>_{\mathcal{B}}|^2=1$ and $|<\downarrow|t>_{\mathcal{B}}|^2=1$, in which case the entire protocol becomes trivial.

This observation does not necessarily invalidate the FR protocol. However, it requires that the first observer's state after an observation is described by the appropriate density matrix.

## 2. Definition of the random measurement process

From the beginning, we do not assume that there is a boundary between "classical" and "quantum" worlds. However, in this section, we shall provide a definition, classical in the sense that the measure adapted to the random process $X_t$ in question is real-valued. We shall generalize this construct to the operator-valued measures in the following section.

To claim that the distinct measurements always produce identical results, these processes must be almost surely identical or coincide everywhere except for the set of the measure null [Kaullenberg02]. This requirement is not practical. At maximum, we can require that the events in the counters coincide only for some mutual set $\{\tau_1<\tau_2, \ldots<\tau_\infty\}$ of the stopping times.

To make this statement more accurate and extendible to the case of quantum measurements, we have to define a measurement process.

**Definition.** The measurement process *M* is a random process adjunct to the random process *X*, if for any (discrete) stopping time τ of *M*.[2]:

$$E[M_t \,|F_\tau^X] = E[f(t) \cdot X_t | F_\tau^M]$$

where *f(t)* is a predictable function. The application of the definition of measurement is based on the following theorem, a loose formulation for which is provided below.

**Theorem 1** (Baxter and Chacon, 1977 [(Baxter77)]).

If the discrete sequences of the stopping times *T(n)*, *U(n)* of the random process *X* do not grow too fast, i.e.

$\lim_{a\to\infty} P(T(n) > a) = 0$ and $\lim_{a\to\infty} P(U(n) > a) = 0$, uniformly in *n* and also $\lim_{n\to\infty} P(T(n) - U(n)) = 0$, then, for any continuous *f(X,Y)*, $\lim_{n\to\infty} P\left[f(X_{T(n)}, X_{U(n)})\right] = 0$.

The meaning of Theorem 1 is very intuitive: if we have two umpires judging the race, if they synchronize the beginning and end of the races, the order of winners will be the same, no matter how imperfect their clocks. The measured speed of each racer can be judged quite differently between them.

**Lemma 1**.

If we have two measurement processes, $M_t^1$, $M_t^2$ adjunct to the same process $X_t$ with the stopping times $\tau_n$, $\xi_n$ satisfying the conditions of Theorem 1, for any convex functional $\rho(X,Y)$, $\forall \varepsilon > 0$

$$\lim_{n\longrightarrow\infty} P\left[\rho\left(M_{\tau_n}^1, M_{\xi_n}^2\right) > \varepsilon\right]=0$$

*Proof*

[2] Note that any adapted random process is trivially measuring for itself. The counting process of a positive process gives a less trivial example. For the finite or countable system of the stopping times $\{\tau_i\}$, we define the auxiliary process by the Equation: $I(t) = \sum_i^{t<T} \theta(t - \tau_i)$, where $T=sup\{\tau_i\}$. The predictable function in our definition will be equal to the $X_{\tau_i}^{-1}$.

By the triangle inequality:

$$\rho\left(M^1_{\tau_n}, M^2_{\xi_n}\right) = \rho\left(M^1_{\tau_n} - M^2_{\tau_n} + M^2_{\xi_n}, M^2_{\xi_n} - M^1_{\tau_n} + M^1_{\xi_n}\right)$$
$$\leq \rho\left(M^1_{\tau_n} - M^1_{\xi_n}, M^2_{\tau_n} - M^2_{\xi_n}\right) + \rho\left(M^1_{\tau_n} - M^2_{\tau_n}, M^1_{\xi_n} - M^2_{\xi_n}\right)$$

Applying uniform convergence in the conditions of Baxter-Chacon's theorem for the first term and choosing ε'=ε/2 and the original Baxter-Chacon theorem for ε' and the second term, we obtain that

$$\lim_{n\to\infty} P\left[\rho\left(M^1_{\tau_n}, M^2_{\xi_n}\right) > \varepsilon\right] \leq$$
$$\lim_{n\to\infty} P\left[\rho\left(M^1_{\tau_n} - M^1_{\xi_n}, M^2_{\tau_n} - M^2_{\xi_n}\right) + \rho\left(M^1_{\tau_n} - M^2_{\tau_n}, M^1_{\xi_n} - M^2_{\xi_n}\right) > \varepsilon\right] = 0$$

Now imagine that the measuring devices $M^1$ and $M^2$ have discrete internal states, which we shall indicate as $i$ and $j$ for $M^1$ and $k$ and $l$ for $M^2$.[3] The notation $i_n$ means that the device $M^1$ was observed in the state $i$ at the $n^{th}$ stopping time. We denote the probability of the $M^1$ transiting $i \to k$ and $M^2$—transiting $j \to l$ at the time τ as $\wp^{jl}_{ik}(\tau)$. We assume a detailed balance between transitions in the same device $\wp^{jl}_{ik}(\tau) = \wp^{lj}_{ki}(\tau)$.

**Theorem 1a**

$$\wp^{jl}_{ik}(\tau) = \wp_{ik}(\tau) \cdot \delta^{jl}_{ik}$$

*Proof*

Note that the convergence in the conditions of Lemma 1 is uniform. Further note that $P[X \in A, Y \in B] = \int_{A\times B} f(x,y)\mu_{X\times Y}(dx,dy)$ is a convex functional for any positive probability density. Consequently, we can apply Baxter-Chacon Theorem 1.

Applying the Bayes formula, we get:

$$\wp^{jl}_{ik}(\tau) = \lim_{n\to\infty} \frac{P\left[M^1_{i_n}|F_\tau\right] \cdot P\left[M^1_{k_n}, M^2_{l_n}\right]}{P\left[M^2_{j_n}|F_\tau\right]} = \lim_{n\to\infty} \frac{P\left[M^1_{i_n}, M^2_{j_n}\right] P\left[M^2_{l_n}|F_\tau\right]}{P\left[M^1_{k_n}|F_\tau\right]} =$$
$$= \wp_{kl}(\tau) \cdot \delta^i_j = \wp_{ij}(\tau) \cdot \delta^k_l$$

[3] See the next section for the justification.

The first set of equalities comes from Theorem 1 and the interchangeability of devices 1 and 2. The second set of equalities comes from Theorem 1 and Lemma 1.

## 3. Building the POVM algebra

For any countable system of the stopping times of the real-valued random process, $\{\tau_i\}$, we can define a counting process by the formula:

$$I(t) = \sum_i^{t<T=\sup\{\tau_i\}} \theta(t-\tau_i) \quad (5)$$

Consequently, for any separable Hilbert space $\mathcal{L}$ with the basis, $\{e_i\}$, we can define the projection-valued measurement process by the formula:

$$M(t) = \begin{cases} I^{-1}(t) \sum_i^{t<T} \hat{P}_i \ \theta(t-\tau_i), \tau_i \le t < T \\ 0, \ 0 \le t < \tau_1 \end{cases} \quad (6)$$

where $\hat{P}_i$ is a projection operator on the basis vector $e_i$. The Equation defines the positive semi-definite system of operators $\tilde{P}_i$:

$$\tilde{P}_i(t) = \begin{cases} M(t), \ \ \tau_i \le t < \tau_{i+1} \\ 0, \ \ otherwise \end{cases} \quad (7)$$

constitutes the partition of the unity operator on $\mathcal{L}$:

$$\hat{I}_{\mathcal{L}} = \sum_i \tilde{P}_i$$

According to the Naimark theorem (Paulsen03), the system of operators $\tilde{P}_i$ can be lifted to the algebra of the operators on the Hilbert space $\mathcal{L}$.

## 4. Measurement in the quantum domain

In this section, we demonstrate that the measurement protocol of the previous section leads to a conventional generalization of the quantum mechanical Born rule.

Heuristically, we can view the mixed state by the wave function depending on two sets of state variables, one of which is the state space of (an unspecified) stochastic process. This stochastic process is described by the family of operator-valued measures $t \rightarrow \mu_t(dQ)$. We identify this stochastic process with the environment/measurement device and take a conditional expectation over (algebra of) states of this stochastic process:

$$\hat{\rho}_t = \int |x, Q >< x', Q)| d\overleftarrow{\mu}_t(dQ) \tag{8}$$

In the symbolic "Equation" (8), the left-handed arrow at the operator-valued measure $\mu(dQ)$ signifies that the operator represented by it acts to the left.

By the construction of the Wigner-Frauchiger-Renner paradox, the algebra of observables can be split into a direct sum of the algebra of observables of two "friends," whom for the referential purpose, we call "Bub" and "Laloë" (Bub2018, Laloe2018):

$$A = A_{\mathcal{B}} \oplus A_L \tag{9}$$

Then, the expectation of Bub's measurement of the observable $\mathcal{A} \in A_{\mathcal{B}}$, which can be represented through the GNS theorem as an operator on the associated Hilbert space (Kadison97), is equal to:

$$E_{\mathcal{L}_\mathcal{B}}\left[\hat{A}_\mathcal{B}\right] = Tr\left[\hat{A}_\mathcal{B} \hat{\rho}_{\mathcal{L}_\mathcal{B}}\right] = \iint \left[\hat{A}_\mathcal{B} |x, Q >< x, Q| \cdot \mu_{\mathcal{L}_B}(dx)\right] \mu_{\mathcal{L}_L}(dQ) \tag{10}$$

After the measurement, the expression in the brackets of Equation (10) undergoes a linear transformation:

$$T: \hat{\rho}_{\mathcal{L}_\mathcal{B}} \rightarrow \hat{\rho}'_{\mathcal{L}_\mathcal{B}}$$

By the Belavkin-Staszewski version of the Radon-Nykodym theorem ([Belavkin86]), the (non-normalized) state of the quantum system $\hat{\rho}'_{\mathcal{L}_\mathcal{B}}$ after the measurement becomes the following:

$$\hat{\rho}'_{\mathcal{L}_\mathcal{B}} = \int V^* \, \tilde{\rho}_Q^{1/2} \, |x, Q >< x', Q| \tilde{\rho}_Q^{\frac{1}{2}} V \cdot \mu_{\mathcal{L}_L}(dQ) \tag{11}$$

In Equation (11), $V$ is a unitary operator, and $\tilde{\rho}$ is a bounded positive self-adjoint operator.

‘ Henceforth, if we take $M = \tilde{\rho}_Q^{\frac{1}{2}} V$, then the state of the quantum system after the measurement will be defined by the standard expression:

$$\hat{\rho}'_{H_1} = \frac{M^* \hat{\rho} M}{Tr_{H_1}(M^* \hat{\rho} M)} \tag{12}$$

Equation (12) is a well-known generalization of the Born rule for probabilities.[4]

## 5. Sharpening of the measurement procedure

The description above encompasses standard measurement procedures in quantum mechanics summarized by the generalized Born rule. As such, it is too general to be useful in most applications. We will now define a measurement process that can be gainfully studied by the current methods.

**Definition**. A *kosher* measurement process is the POVM process defined with an auxiliary triple $\{\mathcal{L}, X_t, \{\tau_i\}\}$, where $\mathcal{L}$ is a Hilbert space, $X_t$ is a random process, and $\{\tau_i\}$ is a countable sequence of the stopping times. The measurement is kosher if it can be represented as a countable sequence:

$$\hat{M}_t = \sum_{i=1} \hat{P}_i(t) I(t) \tag{13}$$

Where

$$\sum_{i=1} \hat{P}_i(t) = 1_{\mathcal{L}} \tag{14}$$

In Equations (13-14), $\hat{P}_i(t)$ are projection operators on the subspaces of the auxiliary Hilbert space $\mathcal{L}$, and $I(t)$ is the counting process for $X_t$, i.e.

$$\left\{ \begin{matrix} I_0 = 0 \\ I(\tau_{i+1}) = I(\tau_i) + 1 \end{matrix} \right\}$$

Not all conceivable or, potentially, even physically realizable measurement processes are necessarily kosher. The notion of the kosher measurement process formalizes measurement devices with a finite number of buttons, dials, and internal memory registers, which are switched on and off for a finite time.

[4] Note that we do not address the “wavefunction collapse." All that we presume, loosely, is that the quantum random walk describes a measurement device. At its stopping time, we observe a distribution conditional on the state of the measurement device, which may or may not make intuitive sense ("superposition of the alive and dead cat").

## 6. Algebra of observables for the "Friends" paradox

For the "Friends" paradox, the algebra of observables splits into a direct sum of the algebras of friends (see Equation (9)). The following diagram can express associated Hilbert spaces.

$$
\begin{array}{ccccccc}
A & \overset{\supseteq}{\leftarrow} & A_1 & \oplus & A_2 & \leftarrow & F_T^* \otimes G_T^* \\
\downarrow & \cong A_1/I_1 & \downarrow & & \downarrow \cong A_2/I_2 & & \uparrow \qquad \uparrow \\
\mathsf{H} & \overset{\supseteq}{\leftarrow} & H_1 & \oplus & H_2 & \leftarrow & F_T \otimes G_T
\end{array}
$$

In Exhibit 1, $F^*_T$, $F_T$, and $G^*_T$, $G_T$ is the Borel algebras for the kosher measurement process between the times [0, T]. Hilbert spaces for Bub and Laloë, $\mathcal{L}_{\mathcal{B}}$, and $\mathcal{L}_L$ are provided by the separate GNS construction (Paulsen03), the maps from direct sum into the "large" Hilbert space and its algebra of observables are trivial. The maps $F^*_T \leftarrow F_T$ and $G^*_T \leftarrow G_T$ are defined correctly because of the Stinespring theorem [(Kadison97, Paulsen03)].

**Lemma 2**. Borel sets $F_T$ and $G_T$ to admit a bijection continuous for all stopping times.

*Proof*

General member of the set $F_T \otimes G_T$ has the form of

$$|\psi >_F \otimes |\psi' >_G = \left\{ \sum_{i,k} a_i \, T_{ik} b_k |i >_{\mathcal{B}} \; |k >_L, \tau_i \leq t < \tau_{i+1} \right\} \tag{15}$$

where $\sum_i a_i T_{ik} = 1$ and $\sum_k b_k T_{ik} = 1$. Therefore,

$$\begin{gathered} \sum_m < m|\psi >_F \otimes |\psi' >_G = \{ \sum_m a_m \, T_{mk} \sum_k b_k \, |k >_L, \tau_i \leq t < \tau_{i+1} \} = \\ \{ \sum_k b_k \, |k >_L, \tau_i \leq t < \tau_{i+1} \} = \left\{ \sum_k < \hat{\hat{P}}_k \, \psi' |k >_L, \tau_i \leq t < \tau_{i+1} \right\} \end{gathered} \tag{16a}$$

and

$$\sum_n |\psi >_F \otimes < n\,|\psi' >_G = \{\sum_n a_i\, T_{in} \sum_n b_n\, |k >_L, \tau_i \leq t < \tau_{i+1} \} = \\ \{\sum_i a_i\, |i >_{\mathcal{B}}, \tau_i \leq t < \tau_{i+1} \} = \left\{\sum_i < \hat{P}_i\, \psi | i >_{\mathcal{B}}, \tau_i \leq t < \tau_{i+1} \right\} \qquad (16b)$$

The last Equations in (16a) and (16b) directly follow from the fact that, by construction, the measurement projections $\hat{P}_i$ and $\widehat{\tilde{P}_k}$ constitute complete systems (see Equation (14)). The following identification provides a required mapping:

$$\sum_i < \hat{\tilde{P}}_k\, \psi' | k >_L\, T_{ki} < i | \widehat{P_\iota} \psi >_{\mathcal{L}}\, | i >_{\mathcal{B}} \leftrightarrow \\ \sum_k < \hat{P}_i\, \psi | i >_{\mathcal{B}}\, T_{ik} < k | \hat{\tilde{P}}_k \psi' >_{\mathcal{L}}\, | k >_L \qquad (17)$$

Because the subspaces /*i*> and /*k*>, by construction, are mutually orthogonal subspaces of $\mathscr{L}_{\mathscr{B}}$ and $\mathscr{L}_L$, respectively, Equation (17) provides a bijection between $F_T$ and $G_T$.

**Theorem 2**. The diagram on Exhibit 1 is correctly defined, and any adjacent set of horizontal and vertical arrows commutes.

*Proof.*

Because of Theorem 1a, the algebras $F_T$ and $G_T$ admit a bijection continuous for all stopping times. The existence of the vertical maps from the operator-valued set into the Hilbert-valued set is a direct consequence of the Stinespring theorem. Henceforth, the maps $F^*_T \leftarrow F_T$ and $G^*_T \leftarrow G_T$ are defined correctly. The only statement to verify is that the horizontal and vertical arrows commute at the left side of the diagram of Exhibit 1. This is true because any state φ on the algebras of the observables $A_{\mathcal{B},L}$, for which $\varphi_i(a_i) = 0$ where $i = \mathcal{B}, L$, and $a_i \in A_i$, belongs to the kernel of the Stinespring map (Kadison97, Paulsen03).

Exhibit 1 already suggests the corrected form of the Frauchiger-Renner postulate Q. Namely, FR postulate that "If $\left\langle \psi \middle| \pi_\xi^{t_0} \middle| \psi \right\rangle = 1$ where $\{\pi_x^t\}$ is a family of Heisenberg operators of S measuring an observable *x*, then $E[x = \xi] = 1$" has to be replaced by the following statement.

Q1. If a family of measurement processes (operator-valued measures) $\{\pi_x^t\}$ defined on an operator algebra being represented as an algebra of operators on a Hilbert space $\mathcal{L}$ stops at $t_0<t$ and

$$Tr_{\mathcal{B}}\left[\hat{\rho}\pi_{\xi}^{t_0}\right] = x,$$

Then

$$E_t\left[x=\xi\middle|F_{t_0}^{\mathcal{B}}\right] = 1.$$

Here $E_t[./F_{t0}]$ is an expectation for the time t given all the information available at $t_0$. Because the spaces $\mathcal{L}_{\mathcal{B}}$ and $\mathcal{L}_L$ do not coincide, they are orthogonal subspaces of $\mathscr{L}$, and there is no contradiction with paradigmatic quantum mechanics in the FR protocol. The equality $E_t\left[x=\xi\middle|F_{t_0}^{\mathcal{B}}\right] = 1$ for Bub's algebra does not entail any definite statement for this expectation with respect to Laloë's algebra $G_t^L$. It can be any number from 0 to 1.

However, this conjecture needs to be clearer and is non-instructive from the physical point of view. We have to refer to the quantum monogamy theorem to demonstrate where the "Friends" protocol breaks down.

## 7. Quantum monogamy and the "Friends" paradox

The main conclusion of this section is that Bub and Laloë are fully entangled. This conclusion can be "rigorously" formulated as Theorem 3.

**Theorem 3**. The "events," i.e., the elements of algebra $F_T \otimes G_T$ from Exhibit 1, are fully entangled.

*Proof.*

First, we start with the algebra of the observables: $F_T^* \otimes G_T^*$. A general-form observable $\hat{\pi}_t \in A$ from the algebra of the observables $A = A_{\mathcal{B}} \oplus A_L$ has the form:

$$\hat{\pi}_t = \{\textstyle\sum_{i\neq k,l\neq m} \wp_{ik}^{lm}(t')\cdot(|i><k|)_{A_{\mathcal{B}}} \otimes {}_{A_L}(|l><m|), \qquad 0 \le t' < t\} \qquad (18)$$

However, according to Theorem 2, a representative member of the algebra $F_T^* \otimes G_T^*$ is much more restricted:

$$\widehat{\pi'}_t = \{\sum_{i \neq k} \wp_{ik}(t') \cdot (|i>< k|)_{A_{\mathcal{B}}} \otimes {}_{A_L}(|k>< i|), \qquad 0 \leq t' < t\} \tag{19}$$

Take kosher projection into an arbitrary two-dimensional subspace of $\mathcal{L}$, $\mathcal{L}_2^0$. For clarity, we shall denote the basis vectors of $\mathcal{L}_2^0$ as |0> and |1>. The resulting state will have the form of an arbitrary $\tau_n$—the stopping time of the measurement process:

$$E[\widehat{\pi'}_{\tau_n} | (F_T^* \otimes G_T^*) \cap A(\mathcal{L}_2^0)] = \\ \wp_{01}(\tau_n) \cdot (|0>< 1|)_{A_{\mathcal{B}}} \otimes_{A_L} (|1>< 0|) + \wp_{10}(\tau_n) \cdot (|1>< 0|)_{A_{\mathcal{B}}} \otimes_{A_L} (|0>< 1|) \tag{20}$$

In Equation (20), $A(\mathcal{L}_2^0)$ is an algebra of the observables on the state space $\mathcal{L}_2^0$. By the normalization condition for the density matrix:

$$|\wp_{01}(\tau_n)| = |\wp_{10}(\tau_n)| = \frac{1}{2}$$

So, the two-dimensional projection of any observable from the joint set of Bub and Laloë is a Bell state, i.e., fully entangled. Because the states |0> and |1> are arbitrary, the state $\hat{\pi}'$ is maximally entangled at all times of completed measurement (the stopping times $\tau_n$ of the measurement process).

## 8. Complete entanglement means no information about the particle

**Theorem 4**. At full entanglement between Bub and Laloë, the state of the incoming particle/photon is undefined.

*Proof*. Without any loss of generality, we suppose that the incoming particle in Part 1 of FR Protocol was detected in a definite "spin up" state or in a similar state of photon polarization, *particle*=|↑>. For arbitrary stopping time $\tau_n$, by the Coffman-Kundu-Wootters version[5] of the Bell theorem ([Coffman00]), the following inequalities—provided here with some abuse of notation—are true:

[5] An interesting question arises: Can quantum monogamy be extended to infinite-dimensional spaces? The author's answer seems positive for the separable Hilbert space, with the proof outside this manuscript's scope. On the

$$C_M^2(\rho_{p\mathcal{B}}) + C_M^2(\rho_{\mathcal{B}L}) \leq C_M^2(\rho_{p(\mathcal{B}L)})$$
$$C_M^2(\rho_{pL}) + C_M^2(\rho_{\mathcal{B}L}) \leq C_M^2(\rho_{p(\mathcal{B}L)}) \quad (21)$$

In Equation (21), '*p*'—is the particle, '$\mathcal{B}$'—is Bub ($F_{\tau_n}$), and '*L*' is Laloe ($G_{\tau_n}$), i.e., their density matrices conditioned on the appropriate algebra. Using the multipartite expression for the concurrence through the density matrix, $C_M(\rho) = \sqrt{2(1 - Tr(\rho_M^2))}$, and the Theorem 3, we obtain:

$$0 \leq C_M^2(\rho_{p\mathcal{B}}) + 2 - O\left(\frac{1}{N}\right) \leq C_M^2(\rho_{p(\mathcal{B}L)}) < 2$$
$$0 \leq C_M^2(\rho_{pL}) + 2 - O\left(\frac{1}{N}\right) \leq C_M^2(\rho_{p(\mathcal{B}L)}) < 2 \quad (22)$$

Equation (22), for *N*—the dimension of the state space of the experimental labs— $N \to \infty$, can be consistent only if $C_M^2(\rho_{p\mathcal{B}}) = C_M^2(\rho_{pL}) = 0$, i.e., both Bub and Laloë do not receive any information about the particle. This conclusion is in full compliance with the quantum monogamy principle.

**Note 1**. Of course, determination of the polarization state of the particle is possible if Bub and Laloë physically coordinate their measurements.[6] However, this would require the non-factorizable Hamiltonian of the system:

$$H = H_{\mathcal{B}} \otimes 1_L + 1_{\mathcal{B}} \otimes H_L + V_{\mathcal{B}L} \neq H_{\mathcal{B}} \otimes 1_L + 1_{\mathcal{B}} \otimes H_L \quad (23)$$

However, this is inconsistent with the initial factorization of the aggregate Hilbert space on Exhibit 1 into Bub and Laloë's subspaces: $\mathcal{L} = \mathcal{L}_{\mathcal{B}} \oplus \mathcal{L}_L$.

**Note 2**. Equation (21) does not preclude a significant amount of information being obtained by Bub and Laloë about the particle, similar to the wave-and-particle duality determination proposed by Wootters and Zurek ([Wootters79]).

In particular, if $C_M^2(\rho_{p(\mathcal{B}L)}) \cong 2$ and $C_M^2(\rho_{p\mathcal{B}}) = 1$, Equation (21) limits information obtained by Laloë only by $C_M^2(\rho_{pL}) \leq 1 - O\left(\frac{1}{N}\right)$.

---

contrary, it is still being determined how this principle can be generalized for the wider variety of state spaces required for quantum mechanics, e.g., Helfand triples.

[6] The author proposed the idea to distinguish between quantum superposition states using Ramsey spectroscopy as early as 1990 [Andreev90].

## Conclusion

Applying a conventional formalism of quantum mechanics provides the following dilemma for our two friends, Bub and Laloë. Namely, to make coincident observations within their rooms about the particle's spin or polarization requires a complete entanglement of their measurement devices. The complete entanglement of their measurements does not allow Bub to determine the state of the particle.

Conceptually, one can imagine that their rooms are connected with an information cable linked to the shutter. This shutter opens or closes Laloë's window precisely when Bub conducts his experiment. However, it is easy to see that introduction of this shutter smears Laloë's measurements because of the uncertainty principle. Using a naïve form of the uncertainty principle, we must admit that this shutter must operate faster than $\tau \sim l/c$, where $l$ is a characteristic size of Bub's room. Such shutter will necessarily introduce a phase shift corresponding to $\Delta\omega \sim c/l$ into the results of Laloë's measurements. Because the propagation time between the rooms is $\sim L/c$, the shutter must be open for at least $\Delta\tau \geq t$ to guarantee that it catches a quantum from Bub. Bub and Laloë occupy different spaces, which guarantees that $L \geq l$. Henceforth, for the shutter scheme $\Delta\omega \cdot \Delta\tau \geq 1$, any diffraction pattern observed by Laloë would be necessarily smeared.

The occurrence of the speed of light in the above reasoning is not random if one notices that in Wigner & Friend paradox, the quantum states of Wigner and Friend's detectors are always separated by a spacelike interval. Quantum states of a particle and Wigner's detector can be separated by a spacelike or timelike interval, dependent on the finer details of an experimental setup.

Our conclusions do not mean that Bub and Laloë cannot extract significant information about each other's experiments, given only partial entanglement between their setups or very cleverly constructed shutter in the spirit of informational inequality of Wootters and Zurek ([Wooters79]).[7] However, any suggestion that adopting Frauchiger and Renner's protocol makes quantum mechanics incomplete should be tabled in this author's view.

[7] In that case, the shutter setup must correspond to the non-factorizable Hamiltonian of Equation (23).

## Conflict of Interest Statement

The paper was not externally supported, and no conflict of interest exists.

**Appendix**: Naïve Tally of Probabilities: Could Wigner Just Be Wrong?

The "Friends" protocol is reexamined with the consistent—the author does not claim physically correct—application of the textbook rules of quantum mechanics. The protocol's conditions are equivalent to the identity of friends' state spaces. In the case of incomplete entanglement of friends, the outcomes of their experiments can vary significantly between them.

For the combined Hilbert space $\mathscr{L}$ of the two friends, named "Bub" and "Laloë" in the main manuscript, the experiment's prior is (see [Rovelli96]):

$$|\Psi>_{\mathcal{L}}=\frac{\alpha}{\sqrt{2}}|\downarrow>_p\,|0>_{\mathcal{B}}+\frac{\beta}{\sqrt{2}}|\uparrow>_p\,|0>_{\mathcal{B}}+\frac{\tilde{\alpha}}{\sqrt{2}}|\downarrow>_p\,|1>_{\mathcal{B}}+\frac{\tilde{\beta}}{\sqrt{2}}|\uparrow>_p\,|1>_{\mathcal{B}} \qquad (1)$$

where $|\alpha|^2+|\beta|^2=1$ and $|\tilde{\alpha}|^2+|\tilde{\beta}|^2=1$. In Equation (1), subscript $p$ is "particle," and subscript $\mathscr{B}$ is "Bub ."The crucial step in the "Friends" protocol is Bub preparing his instrument in the state $|0>_{\mathcal{B}}$ if the particle he registers has spun down and $|1>_{\mathcal{B}}$ if the particle has spin up.

Let Laloë's measurement device be prepared in the state $|0>_L$, i.e., his counter will respond by "0" on Bub's measuring particle in a spin-down state:

$$_L<\psi| = {}_L<0|$$

Then, the non-normalized state of the particle conditional on the settings of Laloë's detector will be:

$$\begin{gathered}_L<\psi|\Psi>_{\mathcal{L}}=\frac{\alpha}{\sqrt{2}}|\downarrow>_p\,{}_L<0|0>_{\mathcal{B}}+\frac{\beta}{\sqrt{2}}|\uparrow>_p\,{}_L<0|0>_{\mathcal{B}}+\\ \frac{\tilde{\alpha}}{\sqrt{2}}|\downarrow>_p\,{}_L<0|1>_{\mathcal{B}}+\frac{\tilde{\beta}}{\sqrt{2}}|\uparrow>_p\,{}_L<0|1>_{\mathcal{B}}=\end{gathered} \qquad (2)$$

$$\frac{1}{\sqrt{2}}\left(\alpha\rho+\tilde{\alpha}e^{i\phi}\sqrt{1-\rho^2}\right)|\downarrow>_p+\frac{1}{\sqrt{2}}\left(\beta\rho+\tilde{\beta}e^{i\phi}\sqrt{1-\rho^2}\right)|\uparrow>_p$$

In the Equation (2) $\rho={}_L<0|0>_{\mathcal{B}}$ and ${}_L<0|1>_{\mathcal{B}}=e^{i\phi}\sqrt{1-\rho^2}$, where $\phi$ is an undetermined phase. The normalization factor for the state ${}_L<\psi|\Psi>_{\mathcal{L}}$ is equal to:

$$A=\left|\,{}_L<\psi|\Psi>_{\mathcal{L}}\right|^2=\frac{1}{2}\left[1+2\left(|\alpha|\cdot|\tilde{\alpha}|+|\beta|\cdot|\tilde{\beta}|\right)\rho\sqrt{1-\rho^2}\,\cos(\phi)\right] \qquad (3)$$

The normalized posterior state is expressed as follows.

$$|\varphi>_{\mathcal{B}} = \frac{{}_{L}<\psi|\Psi>_{\mathcal{L}}}{\left(\left|\,{}_{L}<\psi|\Psi>_{\mathcal{L}}\right|^{2}\right)^{1/2}} = \frac{\frac{1}{\sqrt{2}}\left(\alpha\rho+\tilde{\alpha}e^{i\phi}\sqrt{1-\rho^{2}}\right)|\downarrow>_{p}+\frac{1}{\sqrt{2}}\left(\beta\rho+\tilde{\beta}e^{i\phi}\sqrt{1-\rho^{2}}\right)|\uparrow>_{p}}{\left[1+2\left(|\alpha|\cdot|\tilde{\alpha}|+|\beta|\cdot|\tilde{\beta}|\right)\rho\sqrt{1-\rho^{2}}\cos(\phi)\right]^{1/2}} \quad (4)$$

Using Born's rule, we find the probability of Bub's observation of the particle in the spin-down state according to Laloë's settings of the prior:

$$P_{\mathcal{B}}(\downarrow) = \left|\,{}_{p}<\downarrow|\varphi>_{\mathcal{B}}\right|^{2} = \frac{|\alpha|^{2}\rho^{2}+2|\alpha|\cdot|\tilde{\alpha}|\rho\sqrt{1-\rho^{2}}\cos(\phi)+|\tilde{\alpha}|^{2}\left(1-\rho^{2}\right)}{\left[1+2\left(|\alpha|\cdot|\tilde{\alpha}|+|\beta|\cdot|\tilde{\beta}|\right)\rho\sqrt{1-\rho^{2}}\cos(\phi)\right]} \quad (5)$$

From Equation (5), we observe that only in the case $\rho = \pm 1$, i.e., complete entanglement of the Bub's and Laloë's state spaces, would the measurements by Bub and Laloë agree:

$$P_{\mathcal{B}}(\downarrow) = |\alpha|^{2}$$

In case ρ=0—the orthogonal settings of their detectors, the probability is:

$$P_{\mathcal{B}}(\downarrow) = |\tilde{\alpha}|^{2}$$

This probability has no semblance to Bub preparing "0" with his instrument.

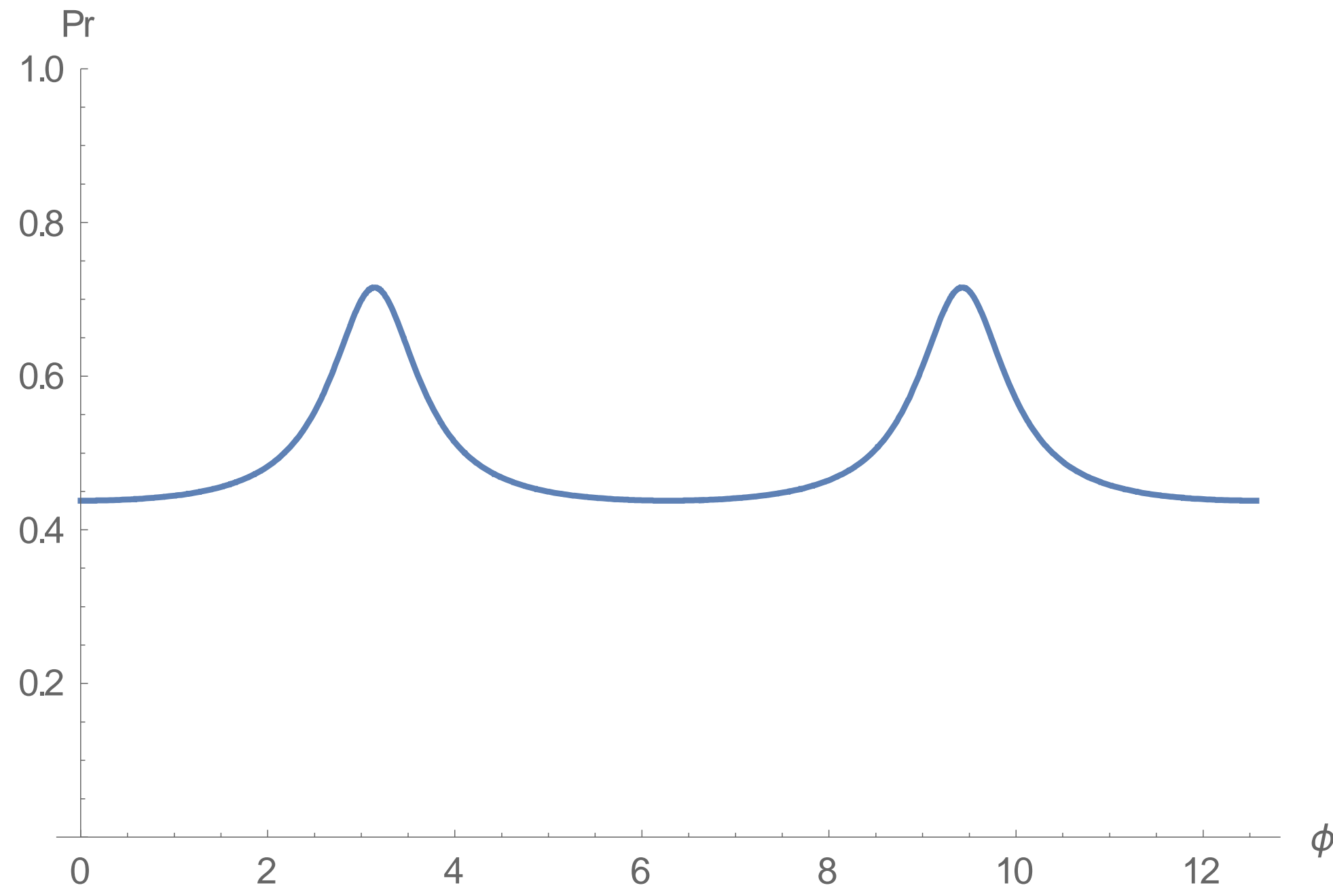

Fig. 1 Measurement results (Equation 5) for the canonic $\alpha = \sqrt{\frac{1}{3}}$ and $\beta = \sqrt{\frac{2}{3}}$ Bub's prior, and $\tilde{\alpha} = \sqrt{\frac{1}{2}}$ and $\tilde{\beta} = \sqrt{\frac{1}{2}}$ for Laloë's prior, indicating his initial lack of information about the spin (polarization) direction. Probabilities measured by Laloë are plotted as a function of deterministic phase shift between bases of their state spaces $\phi$ from 0 to $4\pi$.